\documentclass[
reprint,
showpacs,
amsmath,amssymb,aps,
pra,
]{revtex4-1}

\usepackage{graphicx}% Include figure files
\usepackage{dcolumn}% Align table columns on decimal point
\usepackage{bm}% bold math
\usepackage{hyperref}% add hypertext capabilities
\usepackage[mathlines]{lineno}% Enable numbering of text and display math
\usepackage{xcolor}

\begin{document}

\preprint{}

\title{Direct temporal mode measurement of photon pairs by stimulated emission}% Force line breaks with \\

\author{Xin Chen$^{1}$}
\author{Xiaoying Li$^{2}$}
 \email{xiaoyingli@tju.edu.cn}
\author{Z. Y. Ou$^{1, 2}$}
\email{zou@iupui.edu}
\affiliation{%
$^{1}$Department of Physics, Indiana University-Purdue University Indianapolis, Indianapolis, IN 46202, USA\\
$^{2}$College of Precision Instrument and Opto-Electronics Engineering, Key Laboratory of
Opto-Electronics Information Technology, Ministry of Education, Tianjin University,
Tianjin 300072, P. R. China
}%

%\date{\today}

% To be edited by editor
% \dates{Compiled \today}

%\ociscodes{(270.0270) Quantum Optics; (060.4370) Nonlinear optics, fibers; (190.4380) Nonlinear optics, four wave mixing}

% To be edited by editor
% \doi{\url{http://dx.doi.org/10.1364/optica.XX.XXXXXX}}

\begin{abstract}
It is known that photon pairs  generated from pulse-pumped spontaneous parametric processes can be described by independent temporal modes and form a multi-temporal mode entangled state.
However, the exact form of the temporal modes is not known even though the joint spectral intensity of photon pairs can be measured by the method of stimulated emission tomography. In this paper, we describe a feedback-iteration method which, combined with the stimulated emission method, can give rise to the exact forms of the independent temporal modes for the temporally entangled photon pairs.
\end{abstract}

%\pacs{42.50.Dv, 42.65.Lm, 03.67.Bg}    % PACS, the Physics and Astronomy
%42.50.Ar	Photon statistics and coherence theory
%42.25.Hz	Interference
% arXiv:1805.03397

\maketitle

\section{Introduction}

Pulse-pumped spontaneous parametric processes, because of precise timing provided by the ultra-short pump pulses \cite{zu,tap}, have wide applications in quantum information science such as time-bin entanglement, quantum multi-photon interference of independent sources, heralded single-photon sources. However, the broad bandwidth of the pump field and strict phase matching condition in highly dispersive nonlinear medium lead to complicated spectral correlation in frequency domain.

Fortunately, the issue of complicated spectral correlation was solved in time domain. Law et al. first made a Schmidt decomposition of the joint spectral function and found that the generated two-photon field can be decomposed into a superposition of independent pairs of temporal modes \cite{law}. It was shown later that this mode decomposition can be extended to high gain domain \cite{sil,lvo,guo}. This method significantly simplifies the quantum description for the two-photon fields, leading to multi-dimensional temporal quantum entanglement. Such a temporal mode description was recently extended more generally into field-orthogonal temporal mode analysis of electromagnetic fields and was shown to form a different framework for quantum information \cite{raymer}. Quantum pulse gate technique  through nonlinear interaction processes was developed to distinguish different temporal modes with some success  \cite{sil11,sil14,raymer14,raymer18}.

On the other hand, the specific mode functions of the temporal modes are only revealed by theoretical simulations through the joint spectral function (JSF) of parametric processes \cite{law,sil,lvo,guo}. They can be indirectly obtained through singular value decomposition when the JSF is measured \cite{smith}. But they have never been measured directly until recently when Huo {\it et al} applied a feedback-iteration method to a parametric amplifier operated at high gain regime and found the temporal profiles of the first few temporal modes \cite{huo}. The knowledge of the temporal profiles of the temporal modes then allows the mode-matched homodyne detection of the quantum fields generated by the high gain parametric amplifier to reveal the pair-wise quantum entanglement in continuous variables \cite{huo}. For discrete variables at photon level, the parametric amplifier needs to be operated in the low gain regime for spontaneous emission of photon pairs, as shown by Law {\it et al} \cite{law}. Then, one can use the information of the temporal profile to implement the quantum pulse gates \cite{sil11,sil14,raymer14,raymer18} in temporal mode selection and de-multiplexing.

However, the mode measurement method by Hou {\it et al} \cite{huo} has to rely on the large gain difference among the different modes to eventually lead to the convergence to the mode with highest gain. At the low gain regime of spontaneous emission, the amplifier operates at near unit gain for all modes so there is basically no difference in gain and the method will not lead to a converged shape. One may want to turn up the pump power to push into the high gain regime but it is known that mode structure in parametric processes changes with the pump power at high gain \cite{sipe2,sam}. Thus the method in Ref.\cite{huo} does not work in the low gain regime for spontaneous photon pair generation to reveal the temporal mode structure of the entangled photons discovered by Law et al \cite{law}.

\begin{figure*}
\centering
\includegraphics[width=14cm]{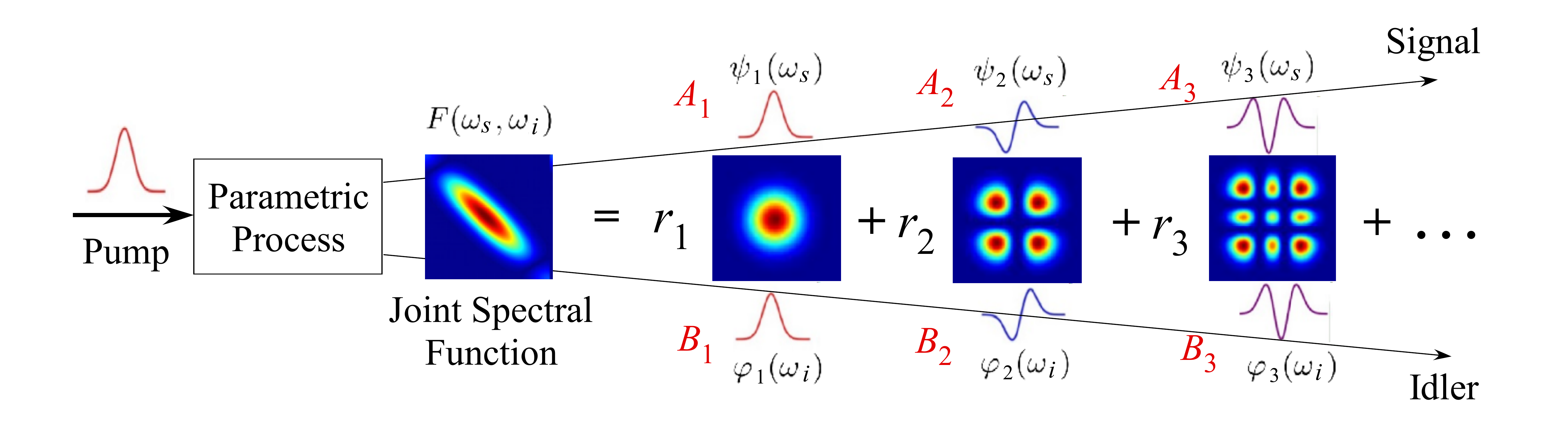}
\caption{Entangled two-photon states consisting of various temporal modes $A_k,B_k (k=1,2,3,...)$ in the signal and idler fields generated from a pulse-pumped parametric process.
}
\label{fig:ill}
\end{figure*}

In this paper, we modify the method by Hou {\it et al.} \cite{huo} and apply it to the low gain regime. In our approach, instead of making measurement on the transmitted and amplified beam, we work on the stimulated emission in the conjugate beam. This is in a way similar to the method of stimulated emission tomography \cite{sipe}. But here we combine the feedback-iteration method in Ref.\cite{huo} and the stimulated emission tomography method in Ref.\cite{sipe} to find the final profiles of modes for both correlated fields in the low gain parametric processes. Since parametric processes at low gain produce two correlated photons, the measured mode functions will be the temporal profiles for the generated photons.

The paper is organized as follows. In Sect.II, we introduce temporal mode analysis of pulse-pumped parametric processes. We then describe in Sect.III our cross-feedback and iteration method for the temporal mode determination at low gain regime and prove the convergence of the iteration. This is based on singular-value decomposition (SVD) of the joint spectral function (JSF). To demonstrate the validity  and the effectiveness of our method described in Sect.III, we present results of numerical simulation in Sect.IV by using input-out relations in parametric processes without the use of the SVD of the JSF. We will also investigate the process of convergence in this section. We conclude with a discussion in Sect.V.

\section{Temporal modes of pulse-pumped parametric processes}

When pumped by an ultra-short pulse, parametric processes, realized via either three-wave or four-wave mixing, produce two correlated fields dubbed ``signal" and ``idler", whose evolution is governed by the unitary operator
\begin{eqnarray}\label{U}
\hat U = \exp\left\{\frac{1}{ i\hbar}\int dt \hat H\right\}
\end{eqnarray}
with \cite{sil,ou-multi}
\begin{eqnarray}\label{H}
\int dt\hat H   = i\hbar \int d \omega_1 d\omega_2F(\omega_1,\omega_2)\hat a_s ^{\dag}(\omega_1)\hat a_i^{\dag}(\omega_2)+ h.c.,~~~~
\end{eqnarray}
where the frequency correlation is described by the joint spectral function (JSF) $F(\omega_1,\omega_2)$ and usually has a complicated form related to the spectral profile of the pump fields and the phase matching conditions. Note that it was realized recently that the unitary operator in the form of Eqs.(\ref{U},\ref{H}) is only approximately correct because $\hat H$ does not commute at different time \cite{sipe2}. See more discussion later at the end of this section.

Fortunately, through the technique of singular value decomposition, it is possible  \cite{sil,lvo,guo} to write the JSF $F(\omega_1,\omega_2)$ in terms of two sets of orthonormal temporal modes $\{\psi_k(\omega_1),\varphi_k(\omega_2)\}$, as shown in Fig.\ref{fig:ill}:
\begin{eqnarray}\label{JSF}
F(\omega_1,\omega_2)= G \sum_k r_k \psi_k(\omega_1) \varphi_k(\omega_2)
\end{eqnarray}
where  $\{r_k\}\ge 0$, satisfying the normalization relation $\sum_k r_k^2=1$, can be arranged in such a way that $r_1\ge r_2\ge ...$ and are the mode numbers. $G>0 $ is a positive dimensionless parameter proportional to the peak amplitudes of the pump fields, nonlinear coefficient and the length of the nonlinear medium, and
\begin{eqnarray}\label{ornorm}
\int d\omega_1\psi^*_k(\omega_1) \psi_{k'}(\omega_1) = \delta_{kk'} = \int d\omega_2\varphi_k^*(\omega_2) \varphi_{k'}(\omega_2).~~~~
\end{eqnarray}
Then Eq.(\ref{H}) can then be rewritten as
\begin{eqnarray}\label{H2}
\frac{1}{ i\hbar}\int dt \hat H = \sum_k G_k \hat A^{\dag}_k \hat B^{\dag}_k - h.c.
\end{eqnarray}
where $G_k\equiv r_kG$, $\hat A_k \equiv \int d\omega_1\psi_k^*(\omega_1)\hat a_s(\omega_1)$ and $\hat B_k \equiv \int d\omega_2\varphi_k^*(\omega_2)\hat a_i(\omega_2)$ are the annihilation operators for the $k$-th modes of the signal and idler fields with respective temporal profiles of $f_k(\tau)\equiv \int d\omega \psi_k(\omega) e^{-i\omega\tau}, g_k(\tau)\equiv \int d\omega \varphi_k(\omega) e^{-i\omega\tau}$.
Notice that different temporal modes ($k$) are decoupled in Eq.(\ref{H2}) so that the input and output relations for the parametric process are \cite{sil,lvo,guo}
\begin{eqnarray}\label{in-out}
\hat A^{out}_k &=& \hat A^{in}_k \cosh G_k +  \hat B^{in\dag}_k \sinh G_k\cr
\hat B^{out}_k &=& \hat B^{in}_k \cosh G_k +  \hat A^{in\dag}_k \sinh G_k,
\end{eqnarray}
which are the relations for a parametric amplifier of amplitude gains $\cosh G_k, \sinh G_k$.
These modes are exactly the super modes studied by Roslund {\it et al.} \cite{fab14} and are independent of each other. Through parametric amplification, these modes are pairwise entangled and form a multi-dimensional quantum entangled states.

For low gain case, $|G_k|\ll 1$ so Eq.(\ref{in-out}) can be approximated as
\begin{eqnarray}\label{in-out2}
\hat A^{out}_k &\approx & \hat A^{in}_k + G_k \hat B^{in\dag}_k  \cr
\hat B^{out}_k &\approx & \hat B^{in}_k + G_k \hat A^{in\dag}_k ,
\end{eqnarray}
or in terms of photon state format, the output state is approximately a two-photon state of the form \cite{law,sil,lvo,guo}
\begin{eqnarray}\label{St}
|\Psi_2\rangle &=& |vac\rangle + \int d \omega_1 d\omega_2F(\omega_1,\omega_2)\hat a_s ^{\dag}(\omega_1)\hat a_i^{\dag}(\omega_2)|vac\rangle\cr
&=& |vac\rangle + \sum_k G_k \hat A^{\dag}_k \hat B^{\dag}_k|vac\rangle\cr
&=& |vac\rangle + G \sum_k r_k |1_{A_k}\rangle_s |1_{B_k}\rangle_i,
\end{eqnarray}
where $|1_{A_k}\rangle_s\equiv \hat A^{\dag}_k|vac\rangle = \int d\omega_1\psi_k(\omega_1)|\omega_1\rangle_s, |1_{B_k}\rangle_i \equiv \hat B^{\dag}_k|vac\rangle = \int d\omega_2\varphi_k(\omega_2)|\omega_2\rangle_i$ are the single-photon states of modes $\hat A_k, \hat B_k$.

It should be noted that because $\hat H(t)$ in general does not commute at different times \cite{sipe2}, Eq.(\ref{U}) is only approximately valid at low gain case. It has been shown \cite{sipe2} that at high gain case, the mode decomposition described in Eq.(\ref{H2}) and the relations in Eq.(\ref{in-out}) are still valid for the two-photon interaction Hamiltonian in Eq.(\ref{H}) but the mode parameters $\{r_k\}$ in gain parameters $G_k= r_k G$ as well as the mode functions $\{\psi_k,\varphi_k\}$ will depend on the pump parameter $G$ \cite{sam}.

Recently, Huo {\it et al.} \cite{huo} applied a feedback-iteration method based on Eq.(\ref{in-out}) to measure explicitly the mode functions $\{\psi_k(\omega_1),\varphi_k(\omega_2)\}$. However, the success of the method relies on the difference in the gain parameters $\cosh G_k$ for different $k$. So, the method fails at low gain case when $\cosh G_k \approx 1$ for all $k$.  In the following, we modify the feedback-iteration method by Huo {\it et al.}\cite{huo} so as to apply to the low gain case to directly measure the mode functions.

\section{Temporal modes determination}
\begin{figure}
\centering
\includegraphics[width=\linewidth]{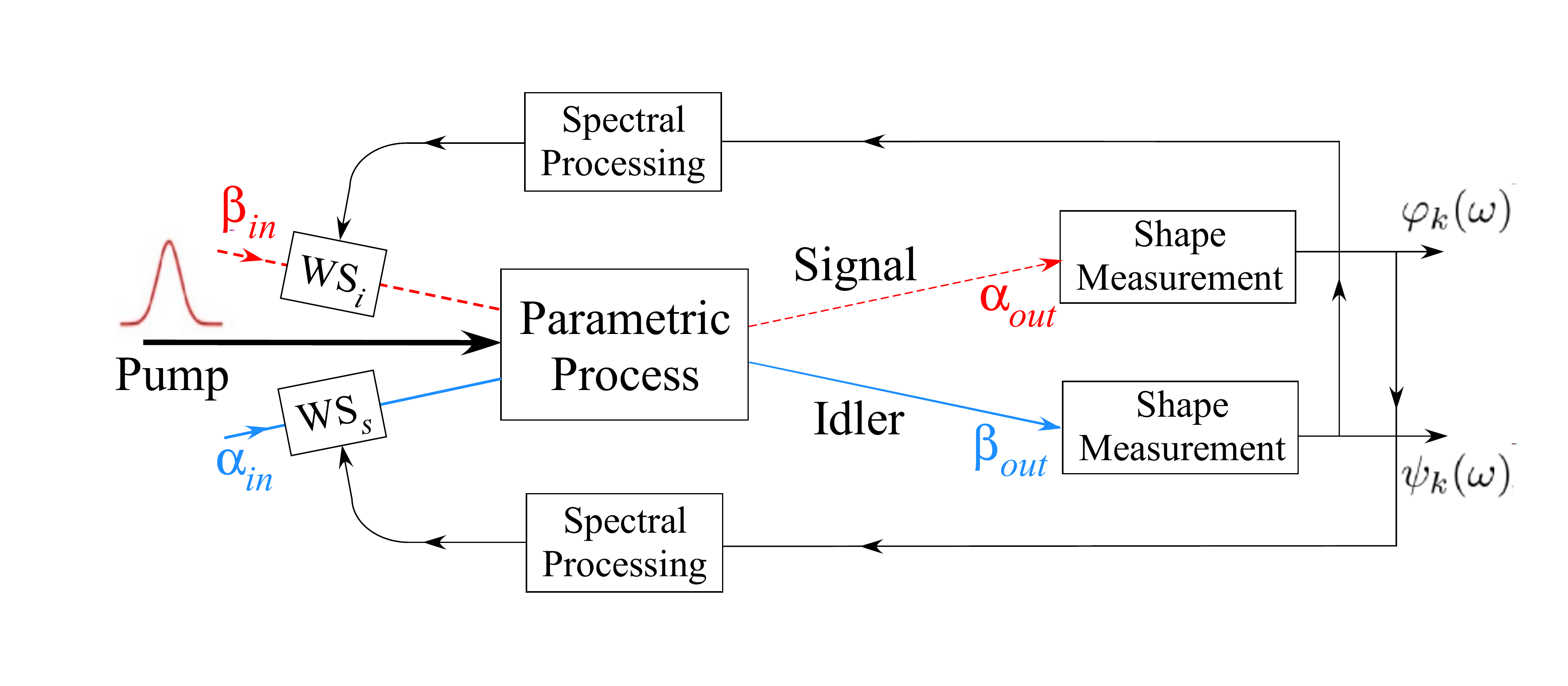}
\caption{Schematic diagram for measuring the mode functions at low gain. WS: wave shaper. The directly related light fields have the same line style (dashed or solid) and color (red or blue)  and the direction of the arrow shows the flow of the iteration.
The converged functions are the outputs that give the measured mode functions $\psi_k(\omega),\varphi_k(\omega)$.}
\label{sch}
\end{figure}

Our procedure to find the mode functions $\psi_k(\omega),\varphi_k(\omega)$ is based on Eq.(\ref{in-out2}).  As shown in Fig.\ref{sch}, we first inject a seed ($\alpha_{in}$) into the signal field and observe the output at idler field ($\beta_{out}$). This is similar to the method of stimulated emission tomography \cite{sipe,lor} but we use the information obtained at the measurement to modify the input seed with wave shapers: with the shape measured at idler ($\beta_{out}$), we then inject this shape of pulse into the idler field and in the meantime observe the output at the signal field ($\alpha_{out}$). Now we have a new shape for the input signal seed. We then alternately inject the seed ($\alpha_{in}$ or $\beta_{in}$) at the signal or idler input based on the measurement result ($\alpha_{out}$ or $\beta_{out}$) and repeat this procedure until steady shapes are observed in both signal and idler fields.

To show the procedure converges, consider a coherent pulse of spectral shape $\alpha_{in}^{(0)}(\omega)$ as the initial injected seed into the signal field $A$. Because of the orthonormality in Eq.(\ref{ornorm}), we can expand it as
\begin{eqnarray}\label{in}
\alpha_{in}^{(0)}(\omega)= \sum_k \xi_k \psi_k(\omega)
\end{eqnarray}
with $\xi_k = \int d\omega \psi_k^*(\omega)\alpha_{in}^{(0)}(\omega)$ as the excitation amplitude for mode $k$. Throughout the paper, we will assume $|\xi_k|^2\gg1$ in order to ignore spontaneous emission in the discussion. Since the gain is nearly 1, the signal output has no information about $G_k$. But it is different for the idler field. Using Eq.(\ref{in-out2}), we find the output at the idler field is approximately
\begin{eqnarray}\label{out}
\beta_{out}^{(1)}(\omega)= \sum_k \xi_k^* G_k \varphi_k(\omega).
\end{eqnarray}
So, the excitations for each mode are modified by $G_k$ but with different coefficients. Now let us exploit this difference in the coefficients: we can measure the output spectral shape $\beta_{out}^{(1)}(\omega)$ at the idler field by using pulse characterization method \cite{wam} and then program an input seed of the shape $\beta_{in}^{(0)}(\omega) = C \beta_{out}^{(1)}(\omega)$ with a wave shaper (WS$_i$). The wave shaper electronic gain constant $C$ can be taken as $C=1/G_1$ to increase the input intensity,  with $G_1$ defined in Eq.(\ref{H2}). At this time, the injection to the signal input is blocked, so the output at the signal field becomes
\begin{eqnarray}\label{out2}
\alpha_{out}^{(1)}(\omega)= \frac{1}{G_1} \sum_k \xi_k G_k^2 \psi_k(\omega).
\end{eqnarray}
Now apply this to another wave shaper (WS$_s$) with the same gain $C=1/G_1$ to produce a new spectral shape for the input seed of the signal field and obtain
\begin{eqnarray}\label{out3}
\alpha_{in}^{(1)}(\omega)&=& C \alpha_{out}^{(1)}(\omega) \cr &=&\frac{1}{G_1^2}\sum_k \xi_k G_k^2 \psi_k(\omega)\cr
&=& \sum_k \xi_k (r_k/r_1)^2 \psi_k(\omega),
\end{eqnarray}
which,  from Eq.(\ref{out}),  leads to the output at the idler:
\begin{eqnarray}\label{out4}
\beta_{out}^{(2)}(\omega)&=& \frac{1}{G_1^2}\sum_k \xi_k^* G_k^3 \varphi_k(\omega)\cr
&=& G_1\sum_k \xi_k^* (r_k/r_1)^3 \varphi_k(\omega),
\end{eqnarray}
Since $r_1> r_2 > ...$, we have $(r_k/r_1)^2 < 1$ for all modes except the first one ($k=1$) and their excitation amplitudes are reduced. We can then iterate the procedure $N$ times and the output field after $N$ iterations becomes
\begin{eqnarray}\label{out5}
\beta_{out}^{(N)}(\omega)&=& G_1\sum_k \xi_k (r_k/r_1)^{2N-1} \varphi_k(\omega)\cr
\alpha_{out}^{(N)}(\omega)&=& G_1\sum_k \xi_k (r_k/r_1)^{2N} \psi_k(\omega).
\end{eqnarray}
With $N$ large enough, $(r_k/r_1)^{2N}\rightarrow 0$ for $k\ne 1$ and we are left with only the first mode: $\alpha_{out}^{(N)}(\omega)\propto \psi_1(\omega)$ and $\beta_{out}^{(N)}(\omega)\propto \varphi_1(\omega)$. This procedure uniquely determines $\psi_1(\omega),\varphi_1(\omega)$ up to a normalization constant.

To obtain the mode function for $k=2$, we need to have an input field that is orthogonal to $\psi_1(\omega)$, that is, $\xi_1=0$. To achieve this, we use the Gram-Schmidt process: with $\psi_1(\omega), \varphi_1(\omega)$ known, we set the input as $\alpha'(\omega) = \alpha(\omega) - \xi_1 \psi_1(\omega)$ or $\beta'(\omega) = \beta(\omega) - \eta_1 \varphi_1(\omega)$ with $\eta_1=\int d\omega \varphi_1^*(\omega)\beta(\omega)$, which gives $\xi_1'=0$ or $\eta'=0$. Then the dominating mode will be $k=2$. To ensure $\xi_1=0$ in the input of each iteration, we perform the orthogonalization step after each measurement of the output. Subsequent modes can be obtained in a similar way but with the orthogonal step changed to $\alpha'(\omega) = \alpha(\omega) - \sum_{i=1}^{k-1}\xi_i \psi_i(\omega)$ or $\beta'(\omega) = \beta(\omega) - \sum_{i=1}^{k-1}\eta_i \varphi_i(\omega)$ for mode $k$.

The argument above is based on the singular value decomposition of the JSF. To demonstrate its validity, we go back to the evolution operator presented in Eq.(\ref{U}) and find the output from the evolution process. Unfortunately, because of the complexity in the JSF, we cannot have an analytical expression so we resort to numerical simulation.

\section{Simulations of temporal mode determination processes}

The evolution operator given in Eq.(\ref{U}) for large pumping power is hard to evaluate \cite{guo13} but
at low pump power for the low gain regime, the dimensionless quantity $G^2\equiv \int d \omega_1 d\omega_2|F(\omega_1,\omega_2)|^2 \ll 1$ and we can expand the exponential in an infinite series and drop the higher order terms. So, the evolution operator can be approximated as \cite{sil,ou-multi}
\begin{eqnarray}\label{U2}
\hat U & = &\exp\left\{\frac{1}{ i\hbar}\int dt \hat H\right\}  \approx  1 + \frac{1}{ i\hbar}\int dt \hat H \cr
& =& 1 + \int d \omega_1 d\omega_2[F(\omega_1,\omega_2)\hat a_s ^{\dag}(\omega_1)\hat a_i^{\dag}(\omega_2)- h.c.].~~~~
\end{eqnarray}
So, the output becomes
\begin{eqnarray}\label{Outs}
\hat a_s^{out}(\omega) &= & \hat U^{\dag} \hat a_s(\omega)\hat U  \cr
& \approx & \hat a_s(\omega) + \int  d\omega_2 F(\omega,\omega_2)\hat a_i^{\dag}(\omega_2),~~~~
\end{eqnarray}
where we used the commutation relation $[\hat a_s(\omega),\hat a_s^{\dag}(\omega_1)]$ $= \delta(\omega-\omega_1)$ and dropped the higher order terms in $F(\omega_1,\omega_2)$.
Similarly,
\begin{eqnarray}\label{Outi}
\hat a_i^{out}(\omega) &= & \hat U^{\dag} \hat a_i(\omega)\hat U  \cr
& \approx & \hat a_i(\omega) + \int d\omega_1 F(\omega_1,\omega)\hat a_s^{\dag}(\omega_1).~~~~
\end{eqnarray}

If we inject a coherent state of $|\{\alpha(\omega)\}\rangle$ at the signal input port but vacuum at the idler port, the expectation value at the idler output will be
\begin{eqnarray}\label{Outi-exp}
\langle \hat a_i^{out}(\omega)\rangle =  \int d\omega_1 F(\omega_1,\omega) \alpha^*(\omega_1)\equiv \beta_{out}(\omega)~~~~
\end{eqnarray}
because the coherent state is independent. Similarly, for an input at the idler port of $|\{\beta(\omega)\}\rangle$, the output at the signal field is
\begin{eqnarray}\label{Outs-exp}
\langle \hat a_s^{out}(\omega)\rangle =  \int d\omega_2 F(\omega,\omega_2) \beta^*(\omega_2)\equiv \alpha_{out}(\omega).~~~~
\end{eqnarray}
Notice that with a singular value decomposition in Eq.(\ref{JSF}) for $F(\omega_1,\omega_2)$ and decomposition of Eq.(\ref{in}) for $\alpha(\omega)$, we recover Eq.(\ref{out}) from Eq.(\ref{Outi-exp}) by using the orthonormal relation in Eq.(\ref{ornorm}).

\begin{figure*}
\centering
\includegraphics[width=\linewidth]{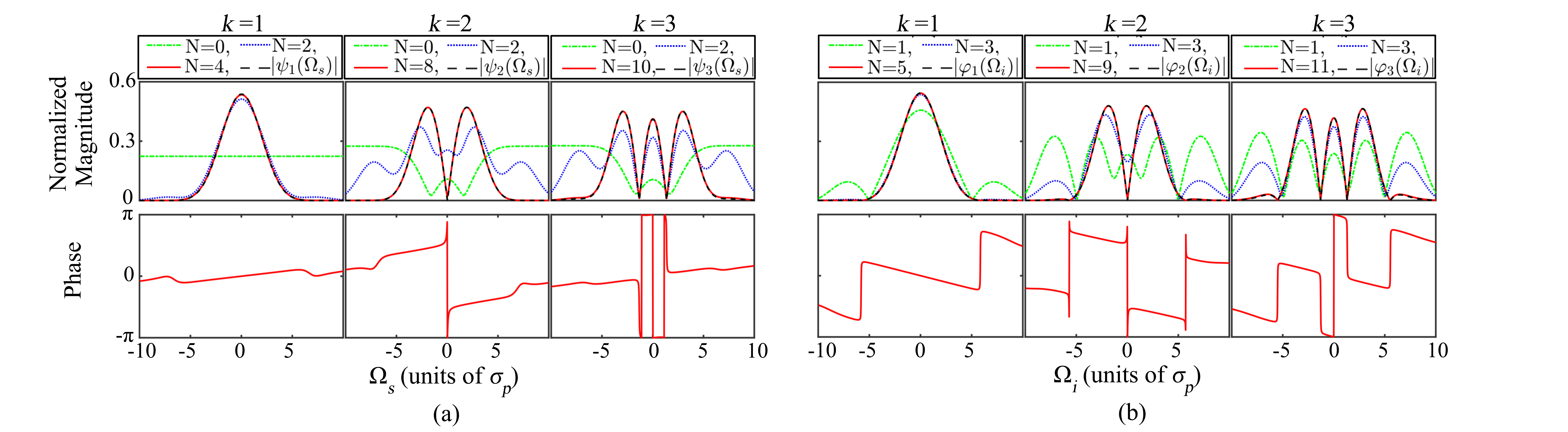}
\caption{Simulated convergent output spectral functions with their magnitudes and phases for the first three modes $k=$ 1, 2, 3 for the JSF given in Eq.(\ref{JSF-sim}). (a) signal field $\psi_k(\Omega_s)$ and (b) idler field $\varphi_k(\Omega_i)$. The green dash-dotted curves are the input spectral functions while the blue dotted and red solid curves are intermediate outputs after the iteration steps indicated in the legends. The black dashed curves are the final outputs.
}
\label{fig3:simu}
\end{figure*}

\begin{figure}
\centering
\includegraphics[width=3 in]{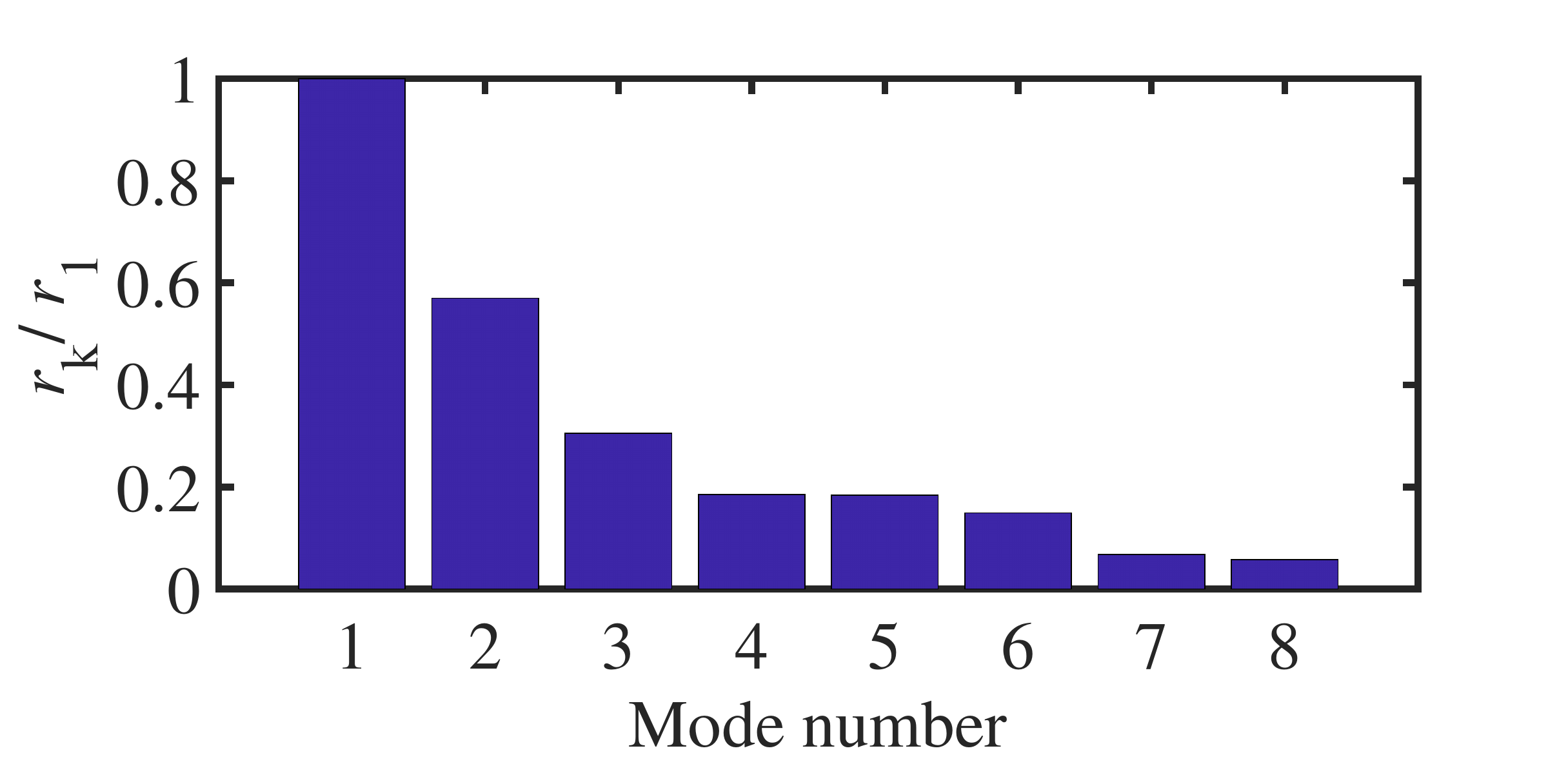}
\caption{Mode number distribution obtained by simulation for the JSF given in Eq.(\ref{JSF-sim}).
}
\label{fig4:simu}
\end{figure}

\begin{figure*}
\centering
\includegraphics[width=\linewidth]{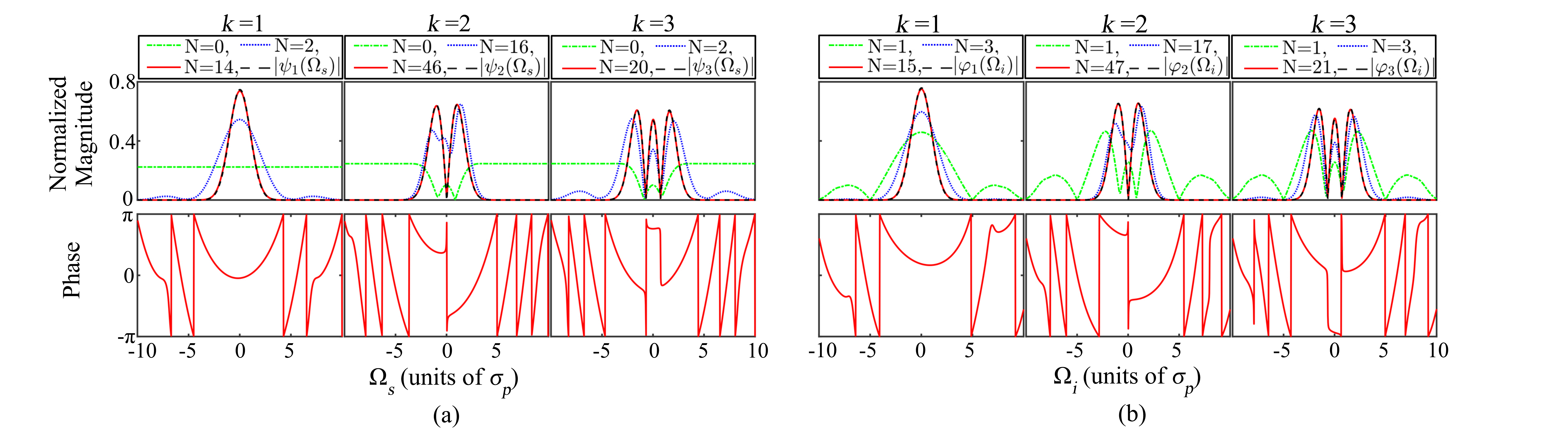}
\caption{Simulated convergent output spectral functions with their magnitudes and phases for the first three modes $k=$ 1, 2, 3 for the JSF given in Eq.(\ref{JSF-sim}) but with a chirped pump phase of $e^{i(\Omega_s+\Omega_i)^2/\sigma_p^2}$. (a) signal field $\psi_k(\Omega_s)$ and (b) idler field $\varphi_k(\Omega_i)$. The green dash-dotted curves are the input spectral functions while the blue dotted and red solid curves are intermediate outputs after the iteration steps indicated in the legends. The black dashed curves are the final outputs.
}
\label{fig5:simu}
\end{figure*}

However, our simulation is based on Eqs.(\ref{Outi-exp}, \ref{Outs-exp}) without the knowledge of the decomposition in Eq.(\ref{JSF}). We will use the cross-feedback method discussed in Sect.III to find the converged functions $\beta_{out}^{c}(\omega)$ and $\alpha_{out}^{c}(\omega)$. From Sect.III, we find $\beta_{out}^{c}(\omega) = \varphi_1(\omega)$ and $\alpha_{out}^{c}(\omega)= \psi_1(\omega)$. So, this cross-feedback and iteration method will lead directly to the first order mode functions $\psi_1(\omega), \varphi_1(\omega)$. We can follow the same procedure in Sect.III to find mode functions of other higher orders.

Furthermore, if we choose a mode-independent electronic gain constant $C$ for the wave shaper, from Eq.(\ref{out3}) we find that once a specific eigenfunction, say, $\psi_{k_0}$ is reached, that is, $\xi_k=\delta_{k,k_0}$, the ratio between next two outputs in the procedure is simply
\begin{eqnarray}\label{Outs-ratio}
\frac{\alpha^{(N+1)}_{out}(\omega)}{\alpha^{(N)}_{out}(\omega)} =  C G^2 r_{k_0}^2\propto r_{k_0}^2.
\end{eqnarray}
So, we can determine the mode numbers $\{r_k\}$ up to a normalization constant.

In order to demonstrate the validity of the procedure above, we consider the JSF given in Ref.\cite{guo} where the parametric process is a pulse-pumped four-wave mixing in a dispersion-shifted fiber. With spectrum shifted to the center frequencies $\omega_{s0}, \omega_{i0}$ of signal and idler beams by defining $\Omega_{s,i} \equiv \omega_{s,i}-\omega_{s0,i0}$, the JSF has the specific form of
\begin{eqnarray}
\label{JSF-sim}
F(\Omega_s,\Omega_i)&&= F  \exp\left \{ -\frac{(\Omega_s+\Omega_i)^2}{4\sigma_p^2}\right \} \cr
 &&~~~~~~~~\times \exp{\left(\frac{-{i \Delta k L}}{2}\right)} {\rm sinc}\left(\frac{\Delta k L}{2}\right).~~~~
\end{eqnarray}
Here $F$ is some constant proportional to the amplitudes of the pump fields and nonlinear coefficient, $\sigma_p$ is the bandwidth of the pump field, $\Delta k L$ is the phase mismatch for fiber length of $L$. For the dispersion-shifted fiber used in Ref.\cite{guo}, it is given by
\begin{equation}
\label{Delta_k}
\frac{\Delta kL}{2} \approx  0.125 \frac{\Omega_s}{\sigma_p}  - 0.075 \frac{\Omega_i}{\sigma_p}.
\end{equation}

\begin{figure}
\centering
\includegraphics[width=3 in]{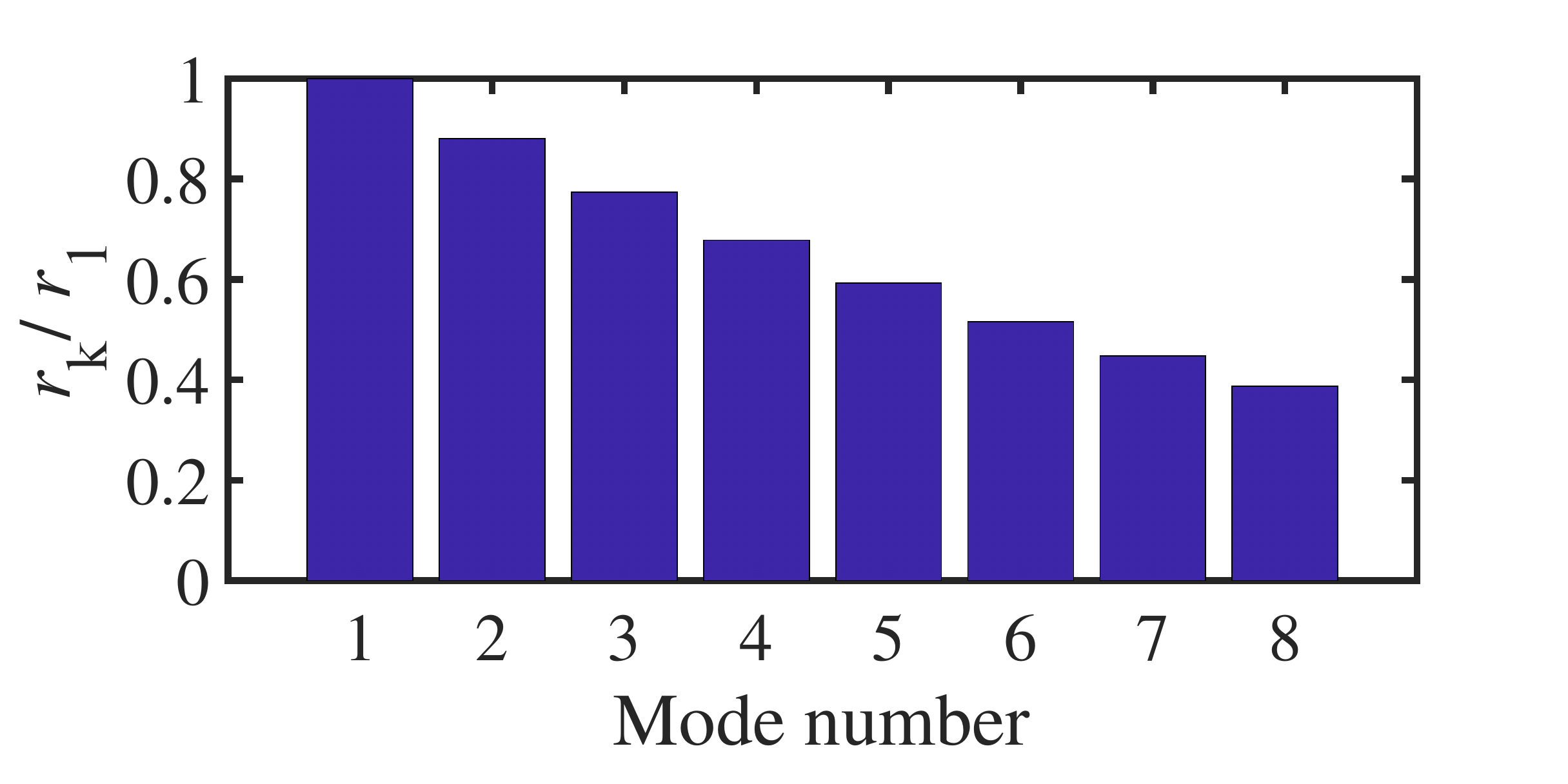}
\caption{Mode number distribution obtained by simulation for the JSF given in Eq.(\ref{JSF-sim})  but with a modified phase of $e^{i(\Omega_s+\Omega_i)^2/\sigma^2_p}$ added due to chirping of the pump field.
}
\label{fig6:simu}
\end{figure}

Our simulation is based on Eq.(\ref{in-out2}), which is derived with the assumption of small $F(\Omega_s,\Omega_i)$ or $F\ll 1$ for low gain case. But because the small $F$ value, the magnitudes of $\beta(\omega)$ and $\alpha(\omega)$ will become progressively decreased as we iterate the process. To maintain the size, we normalize the mode functions $\beta(\omega)$ and $\alpha(\omega)$ after each step of application of Eqs.(\ref{Outi-exp}, \ref{Outs-exp}). So the results are independent of $F$, which is then set to 1 in the simulation. The absolute values and phases of the final converged mode functions of first three orders are shown in Fig.\ref{fig3:simu}(a) for the signal field ($\psi_{1,2,3}(\omega)$) and in Fig.\ref{fig3:simu}(b) for the idler field ($\varphi_{1,2,3}(\omega)$). The green dash-dotted curves are the initial input spectral functions and the black dashed curves are the final output spectral functions. The blue dotted and red solid curves are the output functions in the intermediate steps with the number of iterations shown in the legends. The magnitudes and phases of the mode functions are plotted separately with only final converged phase functions shown. It can be seen that the phase parts vary slowly except the $\pi$-jumps at zeros of the magnitude.  The mode numbers $\{r_k\}$ are plotted in Fig.\ref{fig4:simu} with normalization to $r_1$. It can be seen that the mode functions and the mode numbers are the same as those obtained by the SVD method in Ref.\cite{guo} within the calculation accuracy.

As seen from Fig.\ref{fig3:simu}, the phases of the mode functions vary slowly with the frequency except a jump of $\pi$ at zero points of the functions. This confirms the validity of the approximation of phase as a step function in Ref.\cite{huo}. To see  an example of large phase variation in the mode functions, we add a chirped phase to the spectrum of the pump field resulting in a phase of $e^{i(\Omega_s+\Omega_i)^2/\sigma^2_p}$ to the JSF.  Figure \ref{fig5:simu} shows the magnitudes and phases of the first three mode functions of the signal(a) and idler(b) fields for this case. As can be seen, the phases change rapidly as a function of frequency. Even though the extra chirped phase produces the same joint spectral intensity $|F(\omega_s,\omega_i)|^2$ as that in Eq.(\ref{JSF-sim}), it will change the mode structure as shown in the mode number distribution in Fig.\ref{fig6:simu} as well as the bandwidths of the mode functions in Fig.\ref{fig5:simu}.

To further see the effectiveness of this procedure and the convergence processes, we calculate the ratio of the total output power of the idler to the total input power of the signal for each step, that is,
\begin{eqnarray}\label{ratio}
\left(R^{(2N-1)}\right)^2 \equiv \frac{\int d\omega |\beta^{(N)}_{out}(\omega)|^2}{\int d\omega |\alpha^{(N-1)}_{in}(\omega)|^2}
\end{eqnarray}
and similarly, the ratio of the output at the signal to the input at the idler
\begin{eqnarray}\label{ratio-i}
\left(R^{(2N)}\right)^2 \equiv \frac{\int d\omega |\alpha^{(N)}_{out}(\omega)|^2}{\int d\omega |\beta^{(N-1)}_{in}(\omega)|^2},
\end{eqnarray}
where $N=1,2,3,...$.
These ratios can be measured experimentally. Since $\beta^{(N-1)}_{in}(\omega) \propto \beta^{(N)}_{out}(\omega)$ and $\alpha^{(N)}_{in}(\omega) \propto \alpha^{(N)}_{out}(\omega)$, using Eq.(\ref{out5}) and Eqs.(\ref{out},\ref{out2}), we can find for the first mode
\begin{eqnarray}\label{ratio2}
\left(R_1^{(2N-1)}\right)^2 & =& \frac{\sum_{k=1}^{\infty} |\xi_k|^2G_k^2(r_k/r_1)^{4(N-1)}}{\sum_{k=1}^{\infty} |\xi_k|^2(r_k/r_1)^{4(N-1)}}
\end{eqnarray}
and
\begin{eqnarray}\label{ratio3}
\left(R_1^{(2N)}\right)^2 & =& \frac{\sum_{k=1}^{\infty} |\xi_k|^2G_k^2(r_k/r_1)^{4N-2}}{\sum_{k=1}^{\infty} |\xi_k|^2(r_k/r_1)^{4N-2}}.
\end{eqnarray}
or combining the two cases above for $M=2N-1, 2N$, we have
\begin{eqnarray}\label{ratio4}
\left(R_1^{(M)}\right)^2 & =& \frac{\sum_{k=1}^{\infty} |\xi_k|^2G_k^2(r_k/r_1)^{2M-2}}{\sum_{k=1}^{\infty} |\xi_k|^2(r_k/r_1)^{2M-2}}
\cr\cr &\rightarrow & G^2 r^2_{1} ~~{\rm for}~~ M\rightarrow \infty.
\end{eqnarray}
$M$ is now the overall step number. Likewise, for $k_0$-th mode,
\begin{eqnarray}\label{ratio5}
\left(R_{k_0}^{(M)}\right)^2 & =& \frac{\sum_{k=k_0}^{\infty} |\xi_k|^2G_k^2(r_k/r_{k_0})^{2M-2}}{\sum_{k=k_0}^{\infty} |\xi_k|^2(r_k/r_{k_0})^{2M-2}} \cr\cr &\rightarrow & G^2 r^2_{k_0} ~~{\rm for}~~ M\rightarrow \infty.
\end{eqnarray}

\begin{figure}
\centering
\includegraphics[width=3.2in]{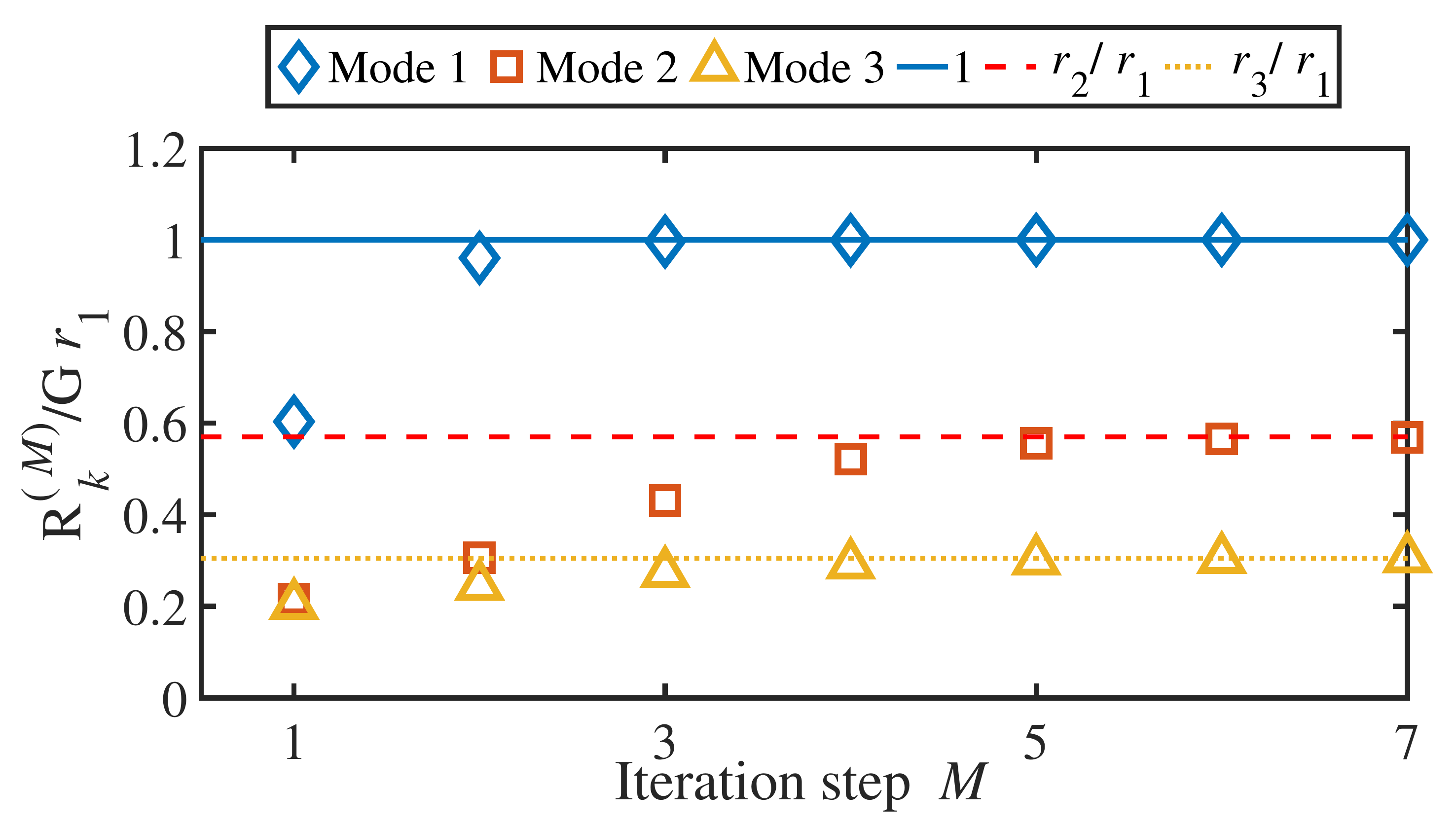}
\caption{Normalized ratio $R_k^{(M)}/Gr_1 (k=1,2,3)$ as a function of iteration step $M$. The straight lines are the limiting values of $r_k/r_1$.
}
\label{fig:modeConv}
\end{figure}

Like $\alpha_{out}^{(N)},\beta_{out}^{(N)}$ in Eq.(\ref{out5}), the convergence of $R_{k_0}^{(M)}$ depends on the ratio $r_k/r_{k_0}$. So quantity  $R_{k_0}^{(M)}$ can represent how the procedure converges as a function of step $M$.
Hence, we calculate $R_{k_0}^{(M)}$ for each iteration step for the $k_0$-th mode ($k_0=1,2,3$) and normalize it to $Gr_1$ for the JSF in Eq.(\ref{JSF-sim}). We plot it as a function of the iteration step numbers in  Fig.\ref{fig:modeConv}. It can be seen that after only a few steps, $R_k^{(M)}$ changes slowly and eventually converges to a final value $r_k/r_1$. So, the rate of convergence is quite good.

\section{Conclusion and Discussion}

We analyze an experimentally implementable method to measure directly the temporal modes for the quantum states generated by pulse-pumped parametric processes. The method is based on the stimulated emission by a trial pulse and relies on a cross-feedback and iteration loop. We demonstrate the convergence of the procedure by numerical simulations for various situation.

Although the simulation is for the low gain case, since the method depends on the difference in the gain coefficients of $G_k$ in Eq.(\ref{in-out2}), it should also work for high gain case where we have different gain coefficients of $\sinh G_k$ in Eq.(\ref{in-out}), except that the mode parameters $\{r_k\}$ in $G_k=r_k G$ are now dependent of $G$. In this case, the mode functions $\{\psi_k,\varphi_k\}$ also depend on the gain \cite{huo,sipe2,sam}.

To check for quantum correlation between different temporal modes for quantum entanglement and orthogonality, as in Ref.\cite{huo}, we need to separate the contributions from different modes, which can be done by homodyne detection in the high gain case \cite{huo} and by quantum pulse gate method in the low gain case \cite{sil11,sil14,raymer14,raymer18}.

The finite spectral response of the shape measurement system may affect the convergence and the outputs. It is equivalent to adding a spectral filter in the iteration loop and thus may change the eigen-modes. In fact, this was observed in the experimental demonstration \cite{huo}: the input and the output may not be the same. This effect will affect higher order modes more than the lower order because higher order modes have wider spectral range.

\begin{acknowledgments}
This work was supported mainly by US National Science Foundation (Grant No. 1806425) and in part by National Natural Science Foundation of China (11527808, 91736105).
\end{acknowledgments}


%\nocite{*}

%\bibliography{REF}% Produces the bibliography via BibTeX.

\end{document}